\documentstyle[12pt]{article}
	\title{Recurrent Clogging and Density Waves in Granular Material
	Flowing through a Narrow Pipe
		\vspace*{2ex}}
	\author{Thorsten P\"oschel}
	\date{\small \today}
\begin{document}
\maketitle
\begin{center}
        preprint HLRZ 67/92 \\
        \vspace*{10ex}
	HLRZ, KFA J\"ulich, Postfach 1913 \\
	D-5170 J\"ulich, Germany  \\
 		\vspace*{1ex}
	and \\
 		\vspace*{1ex}
	Humboldt--Universit\"at zu Berlin \\
	FB Physik, Institut f\"ur Theoretische Physik \\
	Invalidenstra\ss e 42, D-1040 Berlin \\
\end{center}
\vspace*{2ex}
\begin{abstract}
\noindent
We report on density waves in granular material, investigated both
experimentally
and numerically. When granular material falls through a long narrow pipe
one
observes recurrent clogging. The kinetic energy of the falling particles
increases up to a characteristic threshold corresponding to the onset of
recurrent clogging and density waves of no definite wavelength. The
distances
between regions of high density depend strongly on the initial conditions.
They
vary irregularly without any characteristic time and length scale. The
particle--flow was investigated using
2D Molecular Dynamics simulations. Experimental investigations lead to
equivalent results.
\end{abstract}
\vfill

\setcounter{topnumber}{3}
\renewcommand{\topfraction}{1.0}
\newpage
\noindent
The astonishing effects observed in moving granular material have been
subjects
of interest for at least 200 years. Examples of such effects are
the angle of repose \cite{coulomb} \cite{jaeger}
size segregation \cite{haff}--\cite{williams},
heap formation on vibrating media \cite{faraday}--\cite{evesque},
density waves \cite{baxter_89}
and granular flow down inclined surfaces \cite{drake} \cite{savage_89}.
The origin of these effects is
that granular material can behave like a solid or like a liquid, depending
on
its density \cite{reynolds_1}.
Particularly, in recent times
dynamic as well as static behaviour of granular material has been
investigated by many authors experimentally and
theoretically with various
techniques either analytical,
such as thermodynamic \cite{thermod} and hydrodynamic \cite{savage_79}
approaches,
or numerical like cellular automata \cite{baxter_91}, Monte--Carlo
simulation
\cite{devillard} and molecular dynamics \cite{haff} \cite{gallas}
\cite{thompson}.
Even random walk approaches have
been proposed \cite{caram}.
\par
The aim of this Letter is to report the effect that granular material like
sand that falls through a vertical narrow pipe or capillary with a
diameter of
only few particle-diameters changes from a
homogenous to an inhomogenous flow when the kinetic energy $E_k$ of the
falling grains reaches a characteristic threshold. In this state one
observes
recurrent clogging of the particle flow (stick slip motion)
and density waves. We simulated this behaviour using  molecular dynamics
and
demonstrate that the results agree with our experimental observations.
\par
\begin{figure}[ht]
\begin{picture}(13,19.0)
        \put(4.0, 0.0){\line(1,0){6.0}}
        \put(10.0, 0.0){\line(0,1){17.5}}
        \put(10.0, 17.5){\line(-1,0){6.0}}
        \put(4.0, 17.5){\line(0,-1){17.5}}
        \thicklines
        \put(9.5,10.0){\vector(0,-1){1.0}}
\end{picture}
\caption{\it The experiment: fine grained sand with a typical particle diameter
of $D_g=0.18~mm$ flows through a glass capillary tube. Recurrent clogging and
density waves as well as free falling grains are visible. The distances between
the (dark) regions of high density fluctuates significantly. The figure shows
two pipes with different diameters (2~mm and 4~mm, length 1.4~m) and an
enlargement of the thinner one.}
\label{photo}
\vspace{2ex}
\end{figure}
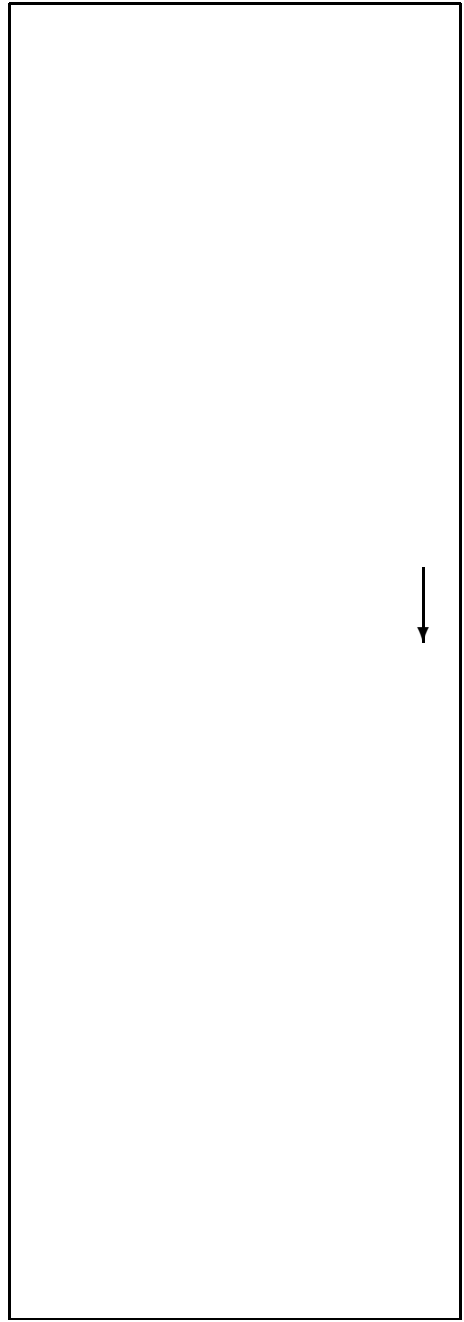

The experiments were carried out in two vertical fixed glass--pipes of
diameters $D_1=2~mm$ and
$D_2=4~mm$ and length $L=1.4~m$. An upper funnel contained enough granular
material to ensure time--independend
initial conditions. During the experiments very fine--grained sand of
typical
particle diameter
$D_g=0.18~mm$ begins to flow homogeneously with the velocity $v=0$ at the
upper end
of the pipe. After a typical distance of approximately $20~cm$ the uniform
flow
becomes unstable and turns into a flow of recurrent clogging. In this
regime
density waves can be observed. Fig.~\ref{photo} shows the pipes with
typical
density distributions. The distances between dark high--density regions vary
irregularly. There is no definite wavelength. The distance between the top of
the tubes and the onset of the stick slip motion as
well as well as for the distances between the plugs depends sensitively on the
humidity of the air and varies significantly from one experiment to the next
one, i.e. depend on initial conditions. According to our observations the
distribution of the distances appears to obey a power law.
\par
\unitlength1.0cm
In fact we have done the simulations first and the reported experiments were
intended to check the accuracy of the results predicted by the numerical
simulations.
\par
According to the experiments reported above we investigated numerically the
flow of spherical-shaped grains with rotational degrees of freedom falling
under gravity
through a pipe of diameter $D$ (fig.~\ref{pipe}) using 2D molecular
dynamics simulations. The length of the pipe considered was much larger then
its diameter $L\gg D$. We investigated systems of particles of equal radii
$r_i=R$ as well as of particles with a Gaussian probability distribution
for the radii with
mean value $R$ (fig.~\ref{snap_micro}). For the flow we assumed periodic
boundary conditions. The inner surface of the pipe is build of smaller spheres
with a diameter $r_s=\frac{2}{3} \cdot R$ of the same material as the grains to
simulate a rough surface. Since our model includes elastic and inelastic
interaction between the grains each other and between the grains and the wall
we
simulated the wall exactly in the same manner as the grains but the particles
the pipe is build of were neither allowed to move nor to rotate.
\par
\unitlength1.0cm
\begin{figure}[ht]
\begin{picture}(12,3.2)
        \thicklines
        \put(1.98,0.9){\line(1,0){9.02}}
        \put(1.98,0.92){\line(1,0){9.02}}
        \put(1.98,0.94){\line(1,0){9.02}}
        \put(1.98,0.96){\line(1,0){9.02}}
        \put(1.98,0.98){\line(1,0){9.02}}
        \put(1.98,1){\line(1,0){9.02}}
        \put(1.98,1.02){\line(1,0){9.02}}
        \put(1.98,1.04){\line(1,0){9.02}}

        \put(1.98,3.06){\line(1,0){9.02}}
        \put(1.98,3.08){\line(1,0){9.02}}
        \put(1.98,3.1){\line(1,0){9.02}}
        \put(1.98,3.12){\line(1,0){9.02}}
        \put(1.98,3.14){\line(1,0){9.02}}
        \put(1.98,3.16){\line(1,0){9.02}}
        \put(1.98,3.18){\line(1,0){9.02}}
        \put(1.98,3.2){\line(1,0){9.02}}

        \put(2.15,1.07){\circle*{0.3}}
        \put(2.55,1.07){\circle*{0.3}}
        \put(2.87,1.07){\circle*{0.3}}
        \put(3.22,1.07){\circle*{0.3}}
        \put(3.70,1.07){\circle*{0.3}}
        \put(4.02,1.07){\circle*{0.3}}
        \put(4.45,1.07){\circle*{0.3}}
        \put(4.77,1.07){\circle*{0.3}}
        \put(5.1,1.07){\circle*{0.3}}
        \put(5.5,1.07){\circle*{0.3}}
        \put(5.95,1.07){\circle*{0.3}}
        \put(6.28,1.07){\circle*{0.3}}
        \put(6.65,1.07){\circle*{0.3}}
        \put(7.1,1.07){\circle*{0.3}}
        \put(7.43,1.07){\circle*{0.3}}
        \put(7.8,1.07){\circle*{0.3}}
        \put(8.1,1.07){\circle*{0.3}}
        \put(8.5,1.07){\circle*{0.3}}
        \put(8.85,1.07){\circle*{0.3}}
        \put(9.3,1.07){\circle*{0.3}}
        \put(9.6,1.07){\circle*{0.3}}
        \put(10.12,1.07){\circle*{0.3}}
        \put(10.52,1.07){\circle*{0.3}}
        \put(10.85,1.07){\circle*{0.3}}

        \put(2.15,3.03){\circle*{0.3}}
        \put(2.55,3.03){\circle*{0.3}}
        \put(2.87,3.03){\circle*{0.3}}
        \put(3.22,3.03){\circle*{0.3}}
        \put(3.70,3.03){\circle*{0.3}}
        \put(4.02,3.03){\circle*{0.3}}
        \put(4.45,3.03){\circle*{0.3}}
        \put(4.77,3.03){\circle*{0.3}}
        \put(5.1,3.03){\circle*{0.3}}
        \put(5.5,3.03){\circle*{0.3}}
        \put(5.95,3.03){\circle*{0.3}}
        \put(6.28,3.03){\circle*{0.3}}
        \put(6.65,3.03){\circle*{0.3}}
        \put(7.1,3.03){\circle*{0.3}}
        \put(7.43,3.03){\circle*{0.3}}
        \put(7.8,3.03){\circle*{0.3}}
        \put(8.1,3.03){\circle*{0.3}}
        \put(8.5,3.03){\circle*{0.3}}
        \put(8.85,3.03){\circle*{0.3}}
        \put(9.3,3.03){\circle*{0.3}}
        \put(9.6,3.03){\circle*{0.3}}
        \put(10.12,3.03){\circle*{0.3}}
        \put(10.52,3.03){\circle*{0.3}}
        \put(10.85,3.03){\circle*{0.3}}

        \thinlines
        \put(1.99,1.0){\line(0,-1){1.0}}
        \put(10.99,1.0){\line(0,-1){1.0}}
        \put(6.0, 0.25){\vector(-1,0){4.0}}
        \put(7.0, 0.25){\vector(1,0){4.0}}
        \put(6.4,0.3){\makebox(0,0)[l]{L}}

        \put(11.0,1.07){\line(1,0){1.0}}
        \put(11.0,3.03){\line(1,0){1.0}}
        \put(11.75, 2.35){\vector(0,1){0.68}}
        \put(11.75, 1.85){\vector(0,-1){0.78}}
        \put(11.6,2.1){\makebox(0,0)[l]{D}}

        \put(8.0,2.1){\vector(1,0){1.0}}
        \put(3.0,2.1){\makebox(0,0)[l]{acceleration due to gravity}}

\end{picture}
\caption{\it The pipe. The rough surface consists of spheres with diameter
	$r_s=\frac{2}{3} \cdot R$ of the same material as the grains.}
\label{pipe}
\vspace{2ex}
\end{figure}
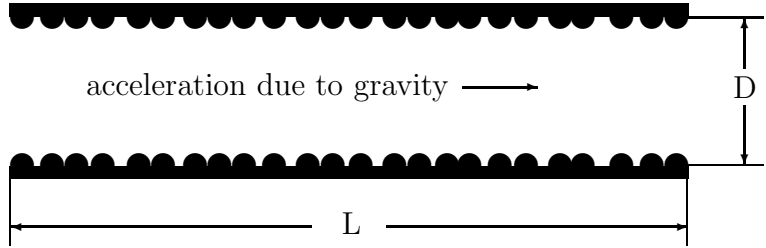

\begin{figure}
\vspace{2ex}
\caption{\it Enlargement of a small part of the pipe filled with grains of
different radii. Gravity acts horizontally from left to right.}
\label{snap_micro}
\vspace{2ex}
\end{figure}
\vspace{2ex}
\noindent
The grains with mass $m_i$ were accelerated due to gravity ($ g~=-9.81~m/s^2 $,
their density $\frac{3\cdot m_i}{4\cdot \pi \cdot {r_i}^3}$
is set to unity):
\begin{equation}
	F_z~=-m_i\cdot g
\end{equation}
\par
\noindent
where $F_z$ is the force acting upon the $i$th particle in the direction of the
axis
of the pipe.
\par
Starting with given initial positions of the particles and randomly chosen
small initial linear and angular velocities the dynamics of
the system was determined by integrating Newtons equation of motion
numerically for each particle for all further time steps. For the integration
we used a sixth order predictor-corrector method for the particle positions and
a fourth order one for the calculation of the particles' angular movement
\cite{predictor}.
\par
In our model particles are allowed to penetrate each other slightly. The force
acting between two particles  $i$ and $j$ is assumed to be zero in case the
particles do not touch each other, i.e. when the distance between their centres
 is
larger then the sum of the particle radii $r_i+r_j$. Otherwise the force
separates into normal and shear force (eq.~\ref{norm+sh_force}).
\par
\begin{equation}
        \vec{F}_{ij} =  \left\{ \begin{array}{cl}
                F_N \cdot \frac{\vec{x}_i - \vec{x}_j}{|\vec{x_i}-\vec{x}_j|} +
                        F_S \cdot  \left({0 \atop 1} ~{-1 \atop 0} \right)
\cdot
                  \frac{\vec{x}_i - \vec{x}_j}{|\vec{x_i}-\vec{x}_j|}
                        & \mbox{if $|\vec{x_i}-\vec{x}_j| < r_i+r_j$} \\[1ex]
                0
                & \mbox{otherwise}
                \end{array}
                \right.
\label{norm+sh_force}
\end{equation}
\par
with
\par
\begin{equation}
        F_N  = k_N \cdot (r_i + r_j - |\vec{x}_i - \vec{x}_j|)^{1.5}~+~\gamma_N
 \cdot m_{eff}(\dot{\vec{x}}_i -
\dot{\vec{x}}_j)
\label{Hertz}
\end{equation}
\par
and
\par
\begin{equation}
        F_S = \min \{- \gamma_S \cdot m_{eff} \cdot v_{rel}~,~ \mu \cdot |F_N|
\}
\label{eq_coulomb}
\end{equation}
\par
where
\par
\begin{equation}
        v_{rel} = (\dot{\vec{x}}_i - \dot{\vec{x}}_j) + R \cdot (\dot{\Omega}_i
 - \dot{\Omega}_j)
\end{equation}
\par
\begin{equation}
        m_{eff} = \frac{m_i \cdot m_j}{m_i + m_j} = \frac{m}{2}
\end{equation}
\vspace{2ex}
\par
\noindent
The  terms $\vec{x}_i$, $\dot{\vec{x}}_i$ and $\dot{\Omega}_i$ stand for the
coordinate, the velocity and the angular velocity of the $i$th particle.
Eq. \ref{Hertz} includes the Hertzian contact force which rises with the
power 1.5 of the penetration depth $r_i + r_j - |\vec{x}_i - \vec{x}_j|$ of two
particles. It is used to mimic 3D behaviour of the grains \cite{landau}. This
ansatz for the force was suggested in~\cite{cundall} and slightly modificated
in~\cite{haff}. Eq.~\ref{eq_coulomb} takes into account, that the maximum
momentum two particles are able to transmit while colliding is determined by
the Coulomb--law \cite{coulomb}. $\mu$ is the Coulomb-coefficient of the
particles.
\par
To discuss the observed phenomena we refer to simulations with $N=600$
particles.  The length of the pipe was $L=666\cdot R$ and its width $D=5\cdot
R$. The shear friction coefficient was $\gamma_S=3\cdot 10^{3}
s^{-1}$, the normal friction coefficient $\gamma_N=3\cdot 10^{3} s^{-1}$, the
material constant $k_N=10^{5} \frac{\mbox{N}}{\mbox{m}^{1.5}}$ and $\mu=0.5$.
For the integration time step was chosen $\Delta t=10^{-5}~s$.
\par
After filling the particles into the pipe they start moving at rest. While
falling they gain
kinetic energy due to the constant acceleration. To avoid infinitely rising
velocity of the particles without colliding they were initialized with a very
small random velocity perpendicular to the axis of the pipe.
At a certain energy the flow does not get faster anymore but the grains are
decelerated very intensively. From now on the system moves with approximately
constant kinetic energy. Fig.~\ref{energy} shows the evolution of the kinetic
energy of the transversal particle movement $E_{k}$ and the energy of the
particle rotation $E_{r}$.
\begin{figure}[ht]
  \vspace{2ex}
  \caption{\it Evolution of the kinetic energy of the transversal particle
  movement and the rotational energy. Since the rotational energy is
  very small the curve is almost indistinguishable of the time-axis. While
flowing
	accelerated, the
  system becomes unstable at 35.000 time steps approximately and transits into
  another flow regime where it reaches a state of relative
  steady kinetic energy ($E\approx 5\cdot 10^9 $for this simulation).}
  \label{energy}
  \vspace{2ex}
\end{figure}
\par
In the case of the low velocity regime the granular material moves through the
pipe
with almost homogeneous particle density. Fig.~\ref{snapshots} displays
snapshots of the pipe each 1,000 time steps. Time increases upwards while
horizontally one sees the evolution of the density wave from the left to the
right. Gravity acts from left
to right. In the early times the
particle-density is approximately homogeneous due to the initial conditions.
During the evolution it gets more and more inhomogeneous and density waves are
observed. As visible at later times the flow is unstable: the regions of high
density can diverge as well as converge while the average velocity remains
approximately the same as shown in fig.~\ref{energy}.
\par
\begin{figure}[h]
\caption{\it Snapshots of the pipe each 1,000 time-steps. Time increases
  upwards. The density fluctuations are not
   equidistant and not stable. They can converge as well as diverge. Dark
  regions correspond to higher densities.}
\label{snapshots}
\vspace{2ex}
\end{figure}
After the transition into the recurrent-plug-flow regime, where the velocity
profile does
not vary with the distance from the centre on the pipe,  the system reaches a
dynamic state
of nearly constant kinetic energy due to the equilibrium between constant
acceleration and energy dissipation by friction. Nevertheless one can observe
small fluctuations of the energy which are aproximately due to the
genesis and vanishing of regions of higher density with time, as visible
 comparing
fig.~\ref{energy}
with fig.~\ref{snapshots}. This behaviour proves that the inhomogenous flow is
unstable and it could be an indication of coexisting metastable states but we
did not check this possibility yet. Our simulations show further that there
are no significant differences between the behaviour of a system with
particles of equal and randomly
distributed radii.
Simulating the system with slightly different initial conditions we got in
principle the same results for the energy evolution and for the structure of
the
density wave but the concrete density wave as it can be observed as dark and
bright regions in fig.~\ref{snapshots} varies significantly. Very small
differences in the initial conditions lead to very different macroscopic
behaviour. In this sense we call the flow chaotically.
\par
There are essential differences between the behaviour of a liquid and
a fluidized granular material: for the case of an incompressible  fluid one
never finds
density waves. Nevertheless there are similarities too: The evolution of the
energy of the system is similar to the evolution of the velocity of a liquid
flowing through a pipe \cite{landau_hyd} while the pressure is increased
gently. At low flow velocity an incompressible viscous fluid
forms a laminar flow (Poisseuille--flow) which gets faster with
rising energy. If the velocity reaches a certain value $u_c$ the
Poiseuille--flow becomes unstable and an inhomogeneous flow regime becomes
stable. For
the case of a fixed pipe geometry and given fluid parameters the critial
velocity of the flow $u_c$ is described by the critical Reynolds--number
\cite{reynolds} $Re_c = \frac{u_c\cdot D}{2\cdot \mu_k} \approx 2,300$ where
$\mu_k$ is the kinematic viscosity of the fluid. Provided there are critical
fluctuations the system turns into the turbulent regime. If the pressure is
increased very gently and the system is not disturbed in another way, however,
it is possible to maintain the Poiseuille--flow at velocities far over
$Re_c$ in the unstable regime. Then a small fluctuation causes the system to
turn to the stable state, i.e. it suddenly lowers the velocity of the flow and
dissipates the energy. As shown for the flow of a granular material
there exist different flow regimes too and the transition between them is due
to a sudden energy dissipation. Hence the transition from low energy
to high energy regime of granular flow in a pipe is similar to
the transition of a laminar fluid-flow into a turbulent one.
\par
Experimental and recent numerical investigations of traffic flows
{\cite{traffic} lead to similar results as those presented here. At a given
average car-velocity the homogeneous traffic flow becomes unstable and traffic
jamming occurs provided there are random perturbations of the velocity of the
cars
which correspond to the collisions of the grains with the wall in our case.
\par
Concluding we state that granular material falling through a narrow pipe
appears
effects far from the behaviour of a liquid. At a certain energy the granular
flow transits from the homogeneous to an inhomogenous flow-regime where
clogging and density waves can be observed. The regions of
high particle density can converge as well as diverge, their distances vary
irregularly. The results of the experiment shown in the beginning agree with
the numerical simulations.
\par
\vspace{5ex}
\par
\noindent
The author thanks J.~Gallas, H.~Herrmann, G.~Ristow and S.~Soko\l owski for
fruitful and stimulating discussions and H.~Puhl for experimental assistance.

\newpage
\listoffigures
\end{document}